\documentclass[12pt,preprint]{aastex}
\usepackage{bbm}
\usepackage{mathrsfs}
\usepackage{amssymb,amsmath,float}
\usepackage{rotating}
\usepackage{color}

\newcommand\cE{\mathcal{E}}

\newcommand\al{\alpha}

\newcommand\de{\delta}
\newcommand\ep{\epsilon}

\renewcommand\th{\theta}

\newcommand\rh{\rho}

\newcommand\om{\omega}

\newcommand\Om{\Omega}

\newcommand\ie{\emph{i.e.}}
\newcommand\eg{\emph{e.g.}}

\newcommand\beq{\begin{equation}}
\newcommand\eeq{\end{equation}}
\newcommand\bea{\begin{eqnarray}}
\newcommand\eea{\end{eqnarray}}
\newcommand\bal{\begin{align}}
\newcommand\eal{\end{align}}

\newcommand\fr{\frac}

\newcommand\ap{\approx}

\newcommand\bk{\bold{k}}

\newcommand\bn{\bold{n}}

\newcommand\br{\bold{r}}

\newcommand\bB{\bold{B}}

\newcommand\bE{\bold{E}}

\renewcommand\bal{\mbox{\boldmath$\alpha$}}

\begin{document}

\title{Are fast radio bursts the birthmark of magnetars?}

\author{Richard Lieu$^1$}

\affil{$^1$Department of Physics, University of Alabama,
Huntsville, AL 35899\\}

\begin{abstract}
A model of fast radio bursts, which enlists young, short period extragalactic magnetars satisfying $B/P >  2 \times 10^{16}$~G~s$^{-1}$ (1~G = 1~statvolt~cm$^{-1}$) as the source, is proposed.  When the parallel component $\bE_\parallel$ of the surface electric field (under the scenario of a vacuum magnetosphere) of such pulsars approaches 5 \% of the critical field $E_c = m_e^2 c^3/(e\hbar)$, in strength, the field can readily decay via the Schwinger mechanism into electron-positron pairs, the back reaction of which causes $\bE_\parallel$ to oscillate on a characteristic timescale smaller than the development of a spark gap.  Thus, under this scenario, the open field line region of the pulsar magnetosphere is controlled by Schwinger pairs, and their large creation and acceleration rates enable the escaping pairs to coherently emit radio waves directly from the polar cap.  The majority of the energy is emitted at frequencies $\lesssim 1$~GHz where the coherent radiation has the highest yield, at a rate large enough to cause the magnetar to lose spin significantly over timescale $\ap$ a few $\times 10^{-3}$~s, the duration of a fast radio burst.  Owing to circumstellar environment of a young magnetar, however, the $\lesssim 1$~GHz radiation is likely to be absorbed or reflected by the overlying matter.  It is shown that the brightness of the remaining (observable) frequencies of $\ap 1$~GHz and above are on par with a typical fast radio burst.  Unless some spin-up mechanism is available to recover the original high rotation rate that triggered the Schwinger mechanism, the fast radio burst will not be repeated again in the same magnetar.

\end{abstract}

\section{Introduction}

Fast radio bursts (FRBs) are a recently discovered and very interesting phenomenon.  The initial discovery of one such event (\cite{lor09}) was followed a few years later by four other similar detections (\cite{tho13}, see also the catalog of FRBs to date in \cite{pet16}).  Their essential properties are a spatial extent consistent with point sources, random transients of brightness $1-10$~Jy lasting a timescale $\sim  1$~ms, typically without repetition (but see \cite{spi16}), and a large line-of-sight dispersion measure that suggests an extragalactic origin. The difficulty in modeling FRBs is the conflict between time and distance scales.  Timescales of order $1$ ms are symptomatic of compact objects like pulsars and stellar mass black holes, these sources are usually too faint to be observable at extragalactic distances.

The discovery of FRBs led us to revisit the subject of compact sources, specifically pulsars, to see if there could be a physical mechanism responsible for their manifestation as extragalactic transients.  Now there exists in the literature an estimate of the brightness of a pulsar (in the radio and other wavelengths).  In the case of an aligned rotator -- the simplified model adopted here -- it is \beq \fr{d\cE}{dt} = \fr{16\pi^4 B^2 R^6}{P^4 c^3} =  9.24 \times 10^{36} \left(\fr{B}{4\times 10^{12}~{\rm G}}\right)^2 \left(\fr{P}{0.1~{\rm s}}\right)^{-4} \left(\fr{R}{10~{\rm km}}\right)^6~{\rm ergs}~{\rm s}^{-1}. \label{dipole} \eeq
Although (\ref{dipole}) agrees with the estimate of \cite{gol69} in their idealized baryonic magnetosphere model, while the more realistic model of \cite{rud75} invokes a spark gap to generate electron-positron pairs, which yielded the more conservative result for the radio luminosity of \beq \fr{d\cE}{dt} = 5.2 \times 10^{32} \left(\fr{B}{10^{12}~{\rm G}}\right)^{6/7} \left(\fr{\rh_c}{10^7~{\rm cm}}\right)^{4/7} \left(\fr{P}{0.1~{\rm s}}\right)^{-15/7}~{\rm ergs~s}^{-1} \label{spark} \eeq where $\rh_c$ is the curvature radius of the surface magnetic field.
In (\ref{spark}), the two parameters that give $d\cE/dt$ its dynamic range are the spin period $P$ and magnetic field $B$, but there are bounds as well.  In particular, to get a large $d\cE/dt$ one needs rapidly rotating and highly magnetized neutron stars.  Although this is an inviting regime because of the discovery of rotationally driven radio magnetars of the spark gap type, \cite{rea12},
the lower limit on $P$ that stems from the need to balance gravity against centrifugal force is $P \gtrsim  0.5$~ms, while magnetic flux conservation during the collapse to a neutron star cannot result in $B$ values far in excess of $10^{15}$~G, the high field end of the magnetar range.

In additional to these constraints, there also exists a correlation between $P$ and $B$, {\it viz.} magnetars tend to have large $P$, of order 1~s or more, \eg~\cite{bel07}; while millisecond pulsars tend to
have $B \lesssim 10^{9}$~G, \eg~\cite{lam05}.  As a consequence of all the above, radio pulsars with $d\cE/dt$ well in excess of (\ref{spark}), \ie~of brightness more approaching the (\ref{dipole}) limit that may enable their detection from extragalactic distances, will probably have to be powered by a very different physical mechanism (operating either temporarily or steadily) from the spark gap model of \cite{rud75}.


The purpose of this paper is indeed to propose a new mechanism of deceleration and rotational energy loss of a pulsar.  If the magnetic field is strong enough to be in the magnetar range, and $P \to 0.1$~s, the parallel electric field $\bE_\parallel$ imposed by boundary conditions upon an idealized vacuum magnetosphere can partially discharge into positron-electron pairs, even though its magnitude is below the Schwinger critical field of \beq E_c = \fr{m_e^2 c^3}{e\hbar} \ap 4.4 \times 10^{13}~{\rm G}.  \eeq  In this regime, one must take account of this extra quantum loss effect that can proceed at a higher rate than the classical limit of (\ref{dipole}).  If an enhancement of $\bE_\parallel$ to such a regime could occur momentarily and briefly,
the result would be a sudden and equally short-lived switch of the pulsar emission scheme, from the low mode of  (\ref{dipole}) to one dominated by the quantum electrodynamic process of vacuum breakdown.  Of course, such short period radio magnetars are probably rare; and it is possible their existence can only be inferred from the much larger extragalactic population of pulsars, via the FRBs.

\section{Schwinger critical electric field}

According to the original calculation of \cite{sch51}, a strong electric field is capable of drawing real e$^+$e$^-$ pairs out of vacuum.
In a pulsar environment where a magnetic field is also present, $\bE$ has vacuum components parallel and perpendicular to $\bB$.  Since the magnitude $E$ is always below $B$, it is possible to transform to a frame in which $\bE \parallel \bB$ by boosting at a non-relativistic speed (the drift velocity), \ie~the results obtained in this frame\footnote{We can ignore the non-inertial effects of this frame, because the pulsar's rotation period is much larger than the timescale of pair creation and annihilation.} is not very different from the discharge of $\bE_\parallel$ in the laboratory frame.  Thus, w.r.t. the $\bE \parallel \bB$ frame, it was shown by \cite{ruf10} that \beq \fr{dn}{dt} = \fr{\al EB}{4\pi^2 \hbar} \coth\left(\fr{\pi B}{E}\right) e^{-\pi E_c/E}~{\rm cm}^{-3}~{\rm s}^{-1}, \label{SchB} \eeq where $\al=e^2/(\hbar c) = 1/137$
is the fine structure constant. Note that in the limit $E \ll \pi B$, which applies throughout this paper, one may ignore the $\coth(\pi B/E)$ factor.

Although the electric fields of interest obey $E \ll B$ and $E\ll E_c$, they are of sufficient strength to accelerate the pairs produced by the discharge, which are initially at rest, in opposite directions.  The field is neutralized as a result.  In detail, the mechanism induces damped oscillations in the field about zero,
the equation of motion of the field may be written by means of Maxwell's 3rd and 4th equations as \beq \ddot E = -4\pi \dot J = -8\pi ce\dot n = -\fr{2\al}{\pi}\fr{ecEB}{\hbar} e^{-\pi E_c/E}~{\rm G}~{s}^{-2} . \label{Edd} \eeq  Note that
as emphasized by \cite{ruf10}, (\ref{Edd}) describes the back reaction of pairs generated by the decay of an electric field on the field itself.  If the pair creation and acceleration rates are high enough to consume the electric field within a cycle, then the oscillation becomes highly damped, and (\ref{Edd}) could at best describe only that fraction of the oscillation cycle up to the moment of full dissipation of the field.



\section{Pair production by direct electric field decay}

To obtain the basic parameters of the electric field oscillations, first observe that it is possible to solve (\ref{Edd}) analytically to the following point: \beq t-t_0 = -\fr{\pi}{2}\left(\fr{\hbar}{\al eBc}\right)^{1/2} \int_{y_0}^y \fr{dy}{(y_0^3 e^{-\pi/y_0} - y^3 e^{-\pi/y})^{1/2}}, \label{Esoln} \eeq where \beq y = \fr{E}{E_c} \label{y} \eeq is the electric field $E(t)$ normalized to the critical field.  The oscillation period may be evaluated by integrating (\ref{Esoln}) down to the first null point $y=0$, {\it viz.} \beq P_{\rm osc} = \fr{2\pi}{\om_{\rm osc}} = 2\pi \left(\fr{\hbar}{\al eBc}\right)^{1/2} \int_{0}^{y_0} \fr{dy}{(y_0^3 e^{-\pi/y_0} - y^3 e^{-\pi/y})^{1/2}}, \label{P} \eeq provided the number of pairs produced per cycle \beq n = \fr{\al E_c B}{2\pi\hbar} \left(\fr{\hbar}{\al eBc}\right)^{1/2} \int_{0}^{y_0} \fr{y e^{-\pi/y} dy}{(y_0^3 e^{-\pi/y_0} - y^3 e^{-\pi/y})^{1/2}} \label{nsoln} \eeq
is small in the sense indicated above (\ref{Edd}).  Assuming the relativistic equation of motion \beq \dot\gamma = \fr{eE}{mc}, \label{eom} \eeq for the pairs, it readily follows that they are accelerated to the Lorentz factor \beq \gamma = \fr{eE_c}{mc} \int_0^{y_0} ydt = \fr{\pi}{2}\left(\fr{\hbar}{\al eBc}\right)^{1/2} \fr{eE_c}{mc} \int_0^{y_0} \fr{ydy}{(y_0^3 e^{-\pi/y_0} - y^3 e^{-\pi/y})^{1/2}}, \label{gamma} \eeq within a quarter cycle of the electric field oscillation.  The results (\ref{P}) to (\ref{gamma}) were represented graphically in \cite{ruf10}.

The acceleration of electrons and positrons depicted in equations (\ref{P}) through (\ref{gamma}) is consistent with that of a large amplitude oscillating electric field.  As a simplified and approximate model, consider an electric field that oscillates as $E(t) = E_0\cos\om_0 t$.  The equation of motion of an electron, $dp/dt = eE$, becomes $\gamma v = eE_0\sin\om_0 t/(m\om_0)$.  In the ultra-relativistic limit of $v \ap c$, this means the pair Lorentz factor $\gamma$ oscillates 90$^\circ$ out of phase with $E$, and with the amplitude \beq \Gamma = \fr{eE_0}{m\om_0 c}. \label{G} \eeq  Indeed, if ones sets $E_0 = y_0 E_c$ and $\om_0 = 2\pi/P_{\rm osc}$, one would find that the resulting $\Gamma$ is on par with (\ref{gamma}).

Now let us get a feel of the oscillation frequency.  As a working example, we set the value of the neutron star magnetic field to $B= 10^{15}$~G (although our conclusion would remain the same if $B$ assumes any value within the magnetar range of $B > E_c$) and the parallel electric field to an initial value $E_0 = 0.045 E_c$, or $y_0 = 0.045$ at the polar field line $\theta = 0$ corresponding to a rotation period $P \ap 0.106$~s via the aligned rotator of \cite{gol69}: \beq E_\parallel = 1.976 \times 10^{12} \left(\fr{P}{0.106~{\rm s}}\right)^{-1} \left(\fr{B}{10^{15}~{\rm G}}\right) \left(\fr{R}{r}\right)^{4} \cos^2 \theta~~{\rm G}. \label{EB} \eeq    A numerical integration of (\ref{P}) then yields \beq P_{\rm osc} = 3.38 \times 10^{-5}~{\rm s,~at}~y_0 = 0.045, \label{Posc045} \eeq while from (\ref{nsoln}) and (\ref{gamma}) the pair number density and Lorentz factor develop to \beq n = 1.57 \times 10^{14}~{\rm cm}^{-3}~{\rm and}~\gamma = 5.35 \times 10^{14},~{\rm at}~y_0 = 0.045 \label{ng045} \eeq respectively, within a quarter oscillation cycle of the field.  Such a number density is on par with the Goldreich-Julian estimate \beq n_{\rm proton} = 7 \times 10^{14} \left(\fr{B_z/10^{15}~{\rm G}}{P/0.1~{\rm s}}\right)~{\rm cm}^{-3}. \label{GJ} \eeq for the protons and electrons pulled out of the stellar surface at comparable speeds (\ie~both speeds being $\ap c$), although the subsequent work of \cite{rud75} presented the difficulties of producing an ionized baryonic magnetosphere and argued for its substitution by pairs, with the number density of primary pair particles at the polar cap surface, \ie~within the `spark gap' region, also commensurate with (\ref{GJ}).

When comparing the \cite{rud75} pairs with the Schwinger pairs proposed here, note that in the former the maximum Lorentz factor is insensitive to $B$ and $P$, {\it viz.}~\beq \gamma_{\rm spark} \lesssim 3 \times 10^6 \left(\fr{B}{10^{12}~{\rm G}}\right)^{1/7} P^{-1/7} \left(\fr{\rh_c}{10^6~{\rm cm}}\right)^{4/7}, \label{gspark} \eeq while in the latter the same is $\propto B/P$ according to (\ref{G}) and (\ref{EB}), and can become $\gg 10^6$. The difference is because the electric field in the spark gap of the former, which has thickness \beq h_{\rm spark} \approx 5 \times 10^3~{\rm cm}, \label{h} \eeq develops and discharges on timescales of 1-10 $\mu$s and never reaches the full length and strength of $\bE_\parallel$ as given by (\ref{EB}) in (even) the region immediately outside the star's surface.  The Schwinger pairs, on the other hand, are born directly from the decay of $\bE_\parallel$ and, as a result, the numbers given after (\ref{EB}) indicate they can reach and surpass the number density of \cite{gol69} and \cite{rud75}, ({\it viz.}~(\ref{GJ}), on equally short timescales or shorter. 
As will be discussed below, this difference between our model and \cite{rud75} is responsible for not only the magnitude of $\gamma$, but also the radio luminosity and its emission region.

But for now, the main point is that provided \beq \fr{B}{P} > 10^{16}~{\rm G~s}^{-1}, \label{limit} \eeq the dynamics of the neutron star magnetosphere in terms of their back reaction on $\bE_\parallel$ are dictated by (\ref{P}) to (\ref{gamma}) and not \cite{rud75}, and this enables the Schwinger pairs to reach much larger $\gamma$ values than the spark gap pairs, even if the spark gap voltages are much higher than the value in (\ref{gspark}) of \cite{rud75}, such as those depicted in Figure 2a of \cite{rea12}.

\section{Power radiated by the Schwinger pairs; radio emission mechanism}

At the field strength of $y_0 = 0.045$ considered here, the rest energy density $2nm_e c^2 = 2.24 \times 10^8$~ergs~cm$^{-3}$ of the pairs is much less than the energy density $E_0^2/(8\pi)= 1.56 \times 10^{23}$~ergs~cm$^{-3}$ of the electric field.  So from this viewpoint the field can oscillate for many cycles without noticeable decay, while the pair density $n$ monotonically increases with time.  Yet the pairs do not remain at rest.   As we saw in the last section, most of the energy of the pairs is kinetic, because they are accelerated by the electric field.


The total electric energy per unit time spent in creating and accelerating the Schwinger pairs along the open magnetic field lines of the polar cap may be estimated  by numerically integrating (\ref{P}), (\ref{nsoln}), and (\ref{gamma}) over the relevant range of $r$, $\theta$, $\phi$, taking into account the dependence of $\bE_\parallel$ upon $r$ and $\theta$, {\it. viz.} (\ref{EB}), with the integration range for $\theta$ limited to $0 \leq \theta \leq \theta_{\rm max}$, where $\theta_{\rm max}$ is given by \beq \sin^2 \theta_{\rm max} = \fr{2\pi R}{Pc} = 1.96 \times 10^{-3} \left(\fr{P}{0.106~{\rm s}}\right)^{-1}. \label{thmax} \eeq and marks the last open field line that crosses the speed-of-light cylinder to transport radiation to infinity.  Moreover, although there is no cutoff height $r$ for $E_\parallel$, the region of sufficient Schwinger breakdown for an appreciable contribution to the total power of the created pairs is found to be limited to \beq r_{\rm max} \approx 0.02R, \label{rmax} \eeq or a scale height $\de R \ap 2 \times 10^4$~cm, a few times larger than the spark gap size of \cite{rud75}, (\ref{h}).
Under the secnario of $y_0 = 0.045$ at $r=R$ and $\theta = 0$,
we find
\beq \fr{d\cE_{\rm pairs}}{dt} = 5.59 \times 10^{41}~{\rm ergs~s}^{-1}, \label{L045} \eeq which is slightly larger than the magnetic dipole radiation rate (\ref{dipole}) for the corresponding values of $B=10^{15}$~G and $P = 0.1$~s.  As $y_0$ increases beyond this point, the pair luminosity rises sharply, see Table 1.


We are ready to discuss the generation of a fast radio burst.  In contrast to \cite{rud75}, wherein the radio signals originated downstream at the light cylinder boundary, the current model can accommodate the much simpler scenario of direct {\it in situ} emission from the polar caps, where the Schwinger discharge is actually occurring, by coherent curvature radiation.  Specifically, the curvature radius for dipole magnetic field lines at the neutron star surface is \beq \rh_c = \fr{R}{3}\fr{1}{\sin\theta}\fr{(1+3\cos^2 \theta)^{3/2}}{1+\cos^2 \theta}, \label{rhc} \eeq see \eg~\cite{smi73}).  For $y_0 \gtrsim 0.095$ and $B=10^{15}$~G, the rotation period is $P \lesssim 0.502$~s from (\ref{EB}), which yields $\th_{\rm max} = 0.0646$ by (\ref{thmax}), hence $\rh_c \lesssim 4.125 \times 10^7$~cm at $\th=\th_{\rm max}/2$.  Now the most conservative estimate of the size $d$ of a region wherein charges are capable of emitting coherently is $d \ap \lambda$, the wavelength of the radiation, \cite{mel92}.  One can see this from the formula of the spectral angular distribution of the intensity emitted by an ensemble of $N$ moving charges in the direction $\hat\bn$ towards the observer for an arrival time of $t$, {\it viz.} \beq \fr{d^2 I}{d\om d\Om} = \fr{q^2\om^2}{16\pi^3 \ep_0 c^3} \Big|\sum_{j=1}^N~\int \hat\bn \times (\hat\bn \times {\bf v_j})~e^{i\om(t- \hat\bn\cdot \br_j/c)} dt\Big|^2, \label{larmor} \eeq that if all the charges are positioned within a radiation wavelength of each other to ensure the phase of each amplitude $e^{-i\om\hat\bn\cdot \br_j/c} = e^{-i\bk\cdot\br_j}$ would differ from the rest by $\ll 2\pi$, and if the velocities ${\bf v_j}$ of each species are essentially the same vector along $\bE_\parallel$ (\ie~bulk speed far exceeds thermal motion, corresponding to negligible collisions) the amplitudes would then add constructively before the modulus squared is taken.  The result is an emitted intensity proportional to $N^2$ rather than $N$.  Hence we designate \beq V_c = \fr{c^3\gamma^2}{\nu^3} = 9.84 \times 10^3 \left(\fr{\nu}{1.4~{\rm GHz}}\right)^{-3} ~{\rm cm}^3 \label{Vc}  \eeq as the coherence volume.

Specific to curvature radiation, the intensity summed over all directions, $dI/d\om$ per emitting charge in units of ergs~s$^{-1}$~Hz$^{-1}$, may be approximated as \beq \fr{dI}{d\om} = 4.67 \times 10^{-23} \rh_c^{-2/3} \om^{1/3}~{\rm ergs~s}^{-1}~{\rm Hz}^{-1}, \label{dIdw} \eeq between $\om=0$ and $\om=3\gamma^3 c/(2\rh_c)$.  If the number density of pairs is $n$, the total emission from the polar cap becomes \bea \fr{d\cE_\gamma}{d\om dt} &=& \fr{dI}{d\om} \times n^2 V_c \times \pi\th_{\rm max}^2 R^2 \de R \notag\\
&=& 2.14 \times 10^{62} \left(\fr{n}{7.60 \times 10^{20}~{\rm cm}^{-3}}\right)^2 \left(\fr{V_c}{9.84 \times 10^3~{\rm cm}^3}\right)
\left(\fr{\th_{\rm max}}{2.4 \times 10^{-3}}\right)^2 \left(\fr{R}{10^6~{\rm cm}}\right)^2 \times \notag\\
& &  \left(\fr{\de R}{2 \times 10^4~{\rm cm}}\right) \left(\fr{\rh_c}{4.125 \times 10^7~{\rm cm}}\right)^{-2/3}~(\om_{\rm min}^{-5/3} - \om_{\rm max}^{-5/3})~{\rm ergs~s}^{-1} \label{Lumin}
\eea
with the effect of coherence taken into account, where the parameters $\th_{\rm max}$ and $\rh_c$ are chosen to suit the $y_0 = 0.095$ scenario for the following reason.  When a young (newly born) magnetar has $y_0$ exceeding this value, then, assuming\footnote{The existence of $\nu_{\rm min}$ is because if $\nu_{\rm min}$ is below $\ap 1$~MHz, the coherence volume $V_c$ would, by (\ref{Vc}), span a thickness $\de R$ in excess of the constraint given by (\ref{rmax}), to include regions where the field is too weak to decay into pairs.} that $\om_{\rm max}$ corresponds to $\nu_{\rm max} = 1.4$~GHz and $\om_{\rm min}$ to $\nu_{\rm min} = 1$~MHz, (\ref{Lumin}) would yield $d\cE_\gamma/dt \ap 10^{51}$~ergs~s$^{-1}$, which from Table 1 is the same as the energy spent on producing and accelerating the pairs.  One can see that in such a regime the energy released by the Schwinger discharge mechanism can all be dissipated into coherent radio emission, which escape along the open field lines.

\section{Conclusion: a model of FRBs as glitch events in fast-spinning young magnetars}

It is now possible to propose an origin of FRB events, as phenomena ensuing from the birth of a magnetar possessing too high an initial spin rate, which leads in turn to the creation of a giant burst of $e^+$ $e^-$ pairs via the Schwinger mechanism.  The energy from this burst is ultimately emitted in the radio wavelengths for reasons given in the previous section, and is drawn from the star's rotation.  In fact, the energy budget is large enough to cause the star to undergo a short episode of decelerating spin-glitch which is the duration of the FRB.

Quantitatively, here is how the numbers work out.  Since the kinetic energy of the neutron star rotation is \beq \cE_{\rm rot} = \tfrac{1}{2} I \Om^2 = 6.27 \times 10^{48} \left(\fr{M_{\rm ns}}{M_\odot}\right) \left(\fr{R}{10^6~{\rm cm}}\right)^2 \left(\fr{P}{0.0502~{\rm s}}\right)^{-2}~{\rm ergs}, \label{nsrot} \eeq the radio emission, which is dominated by radiation below the 1~GHz range, would significantly slow down the rotation in a matter of $5 - 6 \times 10^{-3}$~s, \ie~the timescale of a FRB.  Unless some spin-up mechanism is available to recover the original high rotation rate that triggered the Schwinger mechanism, the fast radio burst will not be repeated again in the same magnetar, because the spin deceleration takes the magnetar outside the regime where the Schwinger mechanism is powerful enough to deliver the luminosity of a FRB.

Moreover, magnetars with the requisite period of $P < 0.1$~s are invariably young, having an upper age limit set by the magnetic dipole radiation rate (the rightmost column of Table 1) and the total available energy (\ref{nsrot}) to a few $\times 10^6$ years.  Such young magnetars are likely to be embedded in the ionized gas of a supernova remnant, and the Schwinger pairs that constitute the aforementioned jet could snowplow into the supernova ejecta to cause a shell of piled-up matter (\cite{che16}); indeed, at least in one case a circumstellar disk was actually observed (\cite{wan06}).
If the $\nu < 1$~GHz frequencies are absorbed or reflected by the shell plasma, as suggested by \cite{kul15}, the only observable radiation would be in the $\ap 1$~GHz, in which case (\ref{Lumin}) would deliver $\ap 5 \times 10^{42}$~ergs in the $\ap 5 \times 10^{-3}$~s duration for the bandwidth of $10$~\% centered at $1.4$~GHz.  This would explain the total radio luminosity of a FRB (note there is no point in including frequency components too much above $1.4$~GHz because, by (\ref{Lumin}) the luminosity has become too low).  It also provides a more substantive basis for the claims of \cite{pop13,kat14,kul15,lyu14,pen15,kat16} of a possible association of FRBs with giant magnetar flares.

In terms of a falsifying test, one possible way is to look for FRBs with duration much longer than 10 ms ($10^{-2}$~s).  They would challenge the proposed model, at least in its simplest form as presented here, because the only way for the magnetar to lose a significant amount of rotational kinetic energy on timescales $\gg 10^{-2}$~s is if the ratio $B/P$ as expressed via $y_0$ falls below 0.09, but in that case the 1 GHz luminosity would also have become too low for observability of the radio burst from an extragalactic distance.  Another test in the same vein would involve discovering more repeating FRBs of the \cite{spi16} type, as again the simplest version of our model cannot account for such a behavior.

\begin{table}[]
\centering
\caption{Properties of Schwinger pairs as a function of the initial ($t=0$) polar surface electric field $y_0$.  The quantities listed in columns 1, 2, 3, 4, and 6 correspond to (\ref{y}), (\ref{P}), (\ref{nsoln}),
(\ref{gamma}), and (\ref{dipole}) respectively, while column 5 is derived numerically as described in the text preceding (\ref{L045}). The surface magnetic field and radius of the neutron star are fixed at $B=10^{15}$~G and $R=10^6$~cm respectively, \ie~any value of $y_0$ can be reached by adjusting the spin period $P$ via (\ref{EB}).}
\vspace{0.5cm}
\label{my-label}
\begin{tabular}{|c|c|c|c|c|c|}
\hline
$y_0$ & $P_{\rm osc}$~(s) & $n$~(cm$^{-3}$) & $\gamma$ & $(d\cE/dt)_{\rm pairs}$ (ergs~s$^{-1}$)  & $(d\cE/dt)_{\rm dipole}$ (ergs~s$^{-1}$) \\ \hline
 0.04  & $2.81 \times 10^{-3}$ & $1.50 \times 10^{12}$ & $3.94 \times 10^{16}$ & $4.79 \times 10^{39}$ & $1.90 \times 10^{41}$ \\ \hline
 0.045 & $3.38 \times 10^{-5}$ & $1.57 \times 10^{14}$  & $5.35 \times 10^{14}$ & $5.59 \times 10^{41}$ &
    $3.05 \times 10^{41}$ \\ \hline
0.055  &  $5.38 \times 10^{-8}$ & $1.19 \times 10^{17}$ & $1.05 \times 10^{12}$ & $6.05 \times 10^{44}$ & $6.81 \times 10^{41}$ \\ \hline
0.085  &  $1.84 \times 10^{-12}$ & $5.37 \times 10^{21}$ & $5.59 \times 10^7$ & $1.09 \times 10^{50}$ & $3.87 \times 10^{42}$ \\ \hline
0.095 & $2.49 \times 10^{-13}$ & $4.40 \times 10^{22}$ & $8.51 \times 10^{6}$ & $1.11 \times 10^{51}$ & $6.05 \times 10^{42}$ \\ \hline
\end{tabular}
\end{table}

\end{document}